\documentclass[a4paper,12pt]{article}

\usepackage[latin1]{inputenc}
\usepackage[centertags]{amsmath}
\usepackage{amsfonts}
\usepackage{amssymb}
\usepackage{amsthm}
\usepackage{newlfont}
\usepackage{braket}
\usepackage{bbm}
\usepackage{graphicx}
\usepackage{color}
\usepackage{dsfont}
\usepackage{setspace}

\theoremstyle{plain}

\theoremstyle{definition}

\theoremstyle{remark}

\numberwithin{equation}{section}

\let\si=\sigma

\newcommand{\bbC}{{\mathbb C}}

\newcommand{\bbN}{{\mathbb N}}

\newcommand{\opunit}{\text{1}\kern-0.22em\text{l}}

\newcommand{\ei}{\epsilon_I}
\newcommand{\eii}{\epsilon_{II}}

\newcommand{\ot}{\tilde{\omega}}

\newcommand{\tr}{\text{Tr} \,}

\begin{document}

\begin{center}
\noindent{\large \bf  Goldstone Bosons in Josephson Junctions } \\

\vspace{15pt}

{\bf Brecht S. Dierckx}\footnote {email:{\tt
brecht.dierckx@fys.kuleuven.be}} , {\bf André F.
Verbeure}\footnote{email: {\tt
andre.verbeure@fys.kuleuven.be}} \\
Instituut voor Theoretische Fysica, K.U.Leuven,
Belgium\\\vspace{10pt}

\end{center}

\vspace{20pt} \footnotesize \noindent {\bf Abstract: } For a
microscopic model of a Josephson junction the normal coordinates
of the two junction Goldstone bosons are constructed and their
dynamical spectrum is computed. The explicit dependence on the
phase difference of the two superconductors is calculated.

\vspace{5pt}
\vspace{5pt}

 \vspace{20pt} \normalsize

\section{Introduction}
\bibliographystyle{plain}
In 1962, Josephson \cite{josephson:1962} predicted a novel
phenomenon in superconductivity, namely when two different
superconductors were brought into close contact. Based on
elementary quantum mechanics, he predicted the existence of a
supercurrent with a peculiar current-voltage dependence. He argued
that there would emerge a current of Cooper pairs which is
proportional to the sine of the phase difference of the order
parameters of both superconductors. The success of this prediction
was immediate when indeed this phenomenon was experimentally
observed one year later \cite{anderson:1963}. It counts as one of
the greatest successes of quantum mechanics in physics and you
will find a chapter on the Josephson effects in almost every
textbook on superconductivity. The increase of knowledge on this
subject in theoretical solid state physics in the following
decades has been tremendous and applications of Josephson
junctions in electronic devices have been developed
\cite{likharev:1986}. Progress in conceiving a microscopic theory
for the Josephson effect in rigorous quantum statistical mechanics
was made when Sewell obtained the Josephson and Meissner effects
in an model independent approach from the assumption of
off-diagonal long range order and local gauge covariance
\cite{sewell:1997}.\\

In \cite{lauwers:2004} we considered a concrete microscopic
quantum model yielding an ab initio and rigorous understanding of
the emerging of a Josephson current, which is computed and
numerically calculated. The model consists of two two-dimensional
superconducting plates having a common one-dimensional contact
surface through which Cooper pairs can tunnel and as such induce a
current. In section 2 we repeat the essentials of the model for
which we construct a non-equilibrium steady state (NESS). One of
the attractive aspects of the construction is that our NESS has
the nice property of having a finite interaction area. We derive
an analytical expression for the current in the case that the
phases of the two superconductors
are not too large. We find back the perfect sine-behavior.\\

Section 3 is devoted to the study of the appearance of Goldstone
bosons in the junction, due to the interaction of the two bulk
superconductors and a direct consequence of the gauge symmetry
breaking. We apply the general result of \cite{michoel2001} where
one finds the general explicit construction of the normal
coordinates of the Goldstone boson as a consequence of the
spontaneous symmetry breaking which appears in our model. The
Goldstone bosons of the bulk superconductors can also be found in
that paper. Here we find two supplementary Goldstone bosons. We
construct their normal coordinates and their dynamics induced by
the micro-dynamics of the model. We consider their dynamics in
diagonal form and analyze its spectrum, again as in section 2 as a
function of the phase difference of the two bulk superconductors.
Here one finds a cosine-behavior.

\section{Micro model for Josephson Junctions}

As said above we consider the model \cite{lauwers:2004} for two
two-dimensional superconducting plates $I$ and $II$ with a common
one-dimensional contact surface (line) through which the Cooper
pairs can travel in order to induce a current.\\
\begin{figure}
\begin{center}
  \includegraphics{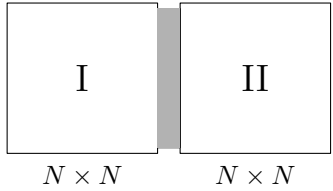}
  \caption{Two superconductors wit a contact surface}
  \end{center}
  \label{fig1}
\end{figure}

The two superconductors are modeled by the strong coupling
BCS-model on a square lattice using the Anderson quasi-spin
formalism and described by the Hamiltonians $H_{i,N}$ with
$i=I,II$

\begin{equation}\label{frham}
H_{i,N}=\sum_{k,l=1}^N \epsilon_{i}\sigma^z(k,l) - \frac
{1}{N}\sum_{k,l,m,n=1}^N
\sigma_{i}^+(k,l)\sigma_{i}^-(m,n),\quad\epsilon_{i}>0
\end{equation}
acting on the Hilbert space $\otimes_{j=1}^{N^2}\bbC_{j}^2$,
 $\sigma_i^{\pm}$ and $\sigma^z=\sigma_i^+\sigma_i^-
-\sigma_i^-\sigma_i^+$ are copies of the Pauli matrices. The
$\sigma^+$ and $\sigma^-$ represent the creation and annihilation
operators of the Cooper pairs of the superconductors; the
$\epsilon$ do represent the kinetic energies of the Cooper
pairs.\\

The junction between the superconductors $I$ and $II$ is modeled
by the interaction

\begin{equation}\label{intham}
V_N=
-\frac{\gamma}{N}\sum_{k_1,k_2=1}^N(\sigma_I^+(k_1,1)\sigma_{II}^-(k_2,1)
 + h.c.),\quad \gamma>0
\end{equation}
which is responsable for the Cooper pair tunneling through the
barrier. A pair at the site $(k_I,1)$ of the first superconductor
can tunnel through the junction and create a pair at the site
$(k_2,1)$ of the second superconductor and vice-versa. The
coupling constant $\gamma$ governs the rate of this process. Note
that only Cooper pairs on the contact surfaces of $I$ and $II$
participate in this process. Remark that only N sites of each
superconductor are interacting with N sites of the other one. The
lattice permutation invariance of the Hamiltonians \eqref{frham}
and the interaction \eqref{intham} make the model given by the
total Hamiltonian of the system
\begin{equation}\label{ham}
H_N=H_{I,N}+H_{II,N}+V_N
\end{equation}
exactly soluble in the thermodynamic limit $N$ tending to infinity
\cite{fannes:1980}.

\subsection {Equilibrium states of the non-interacting
superconductors} For completeness we discuss here the equilibrium
states of the Hamiltonians \eqref{frham}. We treat the first (I)
one explicitly, the second is analogous and obtained by replacing
the index $I$ by the index $II$. Following \cite{fannes:1980}, the
extremal equilibrium states at inverse temperature $\beta_I$ in
the thermodynamic limit are the product states
$\omega_{\varphi_I}$ with the expectation values of all tensor
product observables $X=X_{x_1}\otimes X_{x_2}\otimes ...; \quad
x_1,x_2,...\in \bbN^2$ and all $X_{x_j} \in M_2$(2 by 2 complex
matrices), given by
\begin{equation}\label{frstate}
\omega_{\varphi_I}(X)=\prod_{x\in \mathds{N}^2} \tr
\rho_{\varphi_{I,x}}X_x
\end{equation}
Here $\rho_{\varphi_{I,x}}$ is the x-copy of the 2 by 2 density
matrix $\rho_{\varphi_I}\in M_2$, solution of the selfconsistency
equation
\begin{equation}\label{selfceq}
\rho_{\varphi_I}=\frac{\exp-\beta_I h_{\varphi_I}}{\tr
\exp-\beta_I h_{\varphi_I}}
\end{equation}
with the one-site effective Hamiltonian $h_{\varphi_I}$ given by
\begin{equation}\label{effham}
h_{\varphi_I}=\epsilon_{I}\sigma^z_I-\lambda_I(e^{
i\varphi_I}\sigma_I^- + h.c.), \quad \lambda_I \geq 0
\end{equation}

Clearly the density matrix \ref{selfceq} is also determined by the
equivalent selfconsistency equation for the order parameter
$\lambda_I=|\omega_{\varphi_I}(\sigma_I^-)|$ which by explicit
computation becomes
\begin{equation}
\lambda_I(1-\frac{1}{\mu_I}\tanh\beta_Ik_I)=0, \quad
\mu_I=\sqrt{\epsilon_I^2+\lambda_I^2}
\end{equation}\label{lambda}
Remark that \{$\pm \mu_I$\} constitutes the spectrum of the
effective Hamiltonian $h_{\varphi_I}$ which is independent of the
phase
angle $\varphi_I$.\\
It can readily be seen that \ref{lambda} admits always a solution
$\lambda_I=0$. It yields the \emph{normal phase state} of the
superconductor. For $\varepsilon_I < \frac{1}{2}$ and $\beta_I$
large enough or temperature low enough, there exists also a
solution $\lambda_I\neq{0}$. These solutions yield the
superconducting phase states. The phase $\varphi_I$ can be fixed
freely, yielding an infinite degeneration of the equilibrium
states under the conditions mentioned above.\\The second
superconductor(II) has analogous phase states with phases which
can be chosen independently from the first one.\\In the following
we fix for each superconductor such a superconducting phase state
denoted by $\omega_{I,\varphi_I}$ and $\omega_{II,\varphi_{II}}$
with $\varphi_I\neq{\varphi_{II}}$.\\

\subsection{Non-equilibrium steady state (NESS)}
Now we construct a non-equilibrium but steady state (NESS) for the
total interacting system \ref{ham}. We start from the product
state $\omega=\omega_{I,\varphi_I}\otimes\omega_{II,\varphi_{II}}$
on the system, i.e. the state of the system of the two
superconductors in their respective superconducting phase states
characterized by their phases, inverse temperatures and kinetic
energies. In this state $\omega$ one can compute the global
dynamics yielding a time evolution
\begin{equation}\label{dyn}
\alpha_t(.)=\omega-\lim_N(\exp{itH_N} . \exp{itH_N})
\end{equation}
where $\omega-\lim$ is the weak limit under the state $\omega$.
Now we are ready to look for a state $\tilde{\omega}$ which is
invariant under the dynamics \ref{dyn}.\\
Due to the specific lattice permutation symmetry of the model, it
is natural to choose this state $\tilde{\omega}$ among the product
states \cite{fannes:1980}, i.e.
\begin{equation}\label{state}
\tilde{\omega}_{\varphi_I}(X)= \tr \prod_{x\in
\bbN^2}\tilde{\rho}_{x}X_x
\end{equation}
As the state should be time invariant, it should have the same
lattice invariance as the Hamiltonian \ref{ham}. The symmetry of
the Hamiltonian divides the total system into four parts.
\begin{figure}[h]\label{fig2}
\begin{center}
\includegraphics{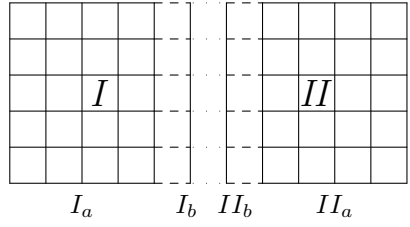}
\caption{Division of the system in four subsystems}
\end{center}
\end{figure}

There are the bulk parts of the superconductors $I$ and $II$ which
we denote by $I_a$ and $II_a$, and there are the contact or
surface parts denoted by $I_b$ and $II_b$. Therefore one can write
the state $\tilde{\omega}$ as a tensor product of four symmetric
product states on their different regions:

\begin{equation}
\tilde{\omega}=\tilde{\omega}_{I_a}\otimes\tilde{\omega}_{I_b}\otimes\tilde{\omega}_{II_b}\otimes\tilde{\omega}_{I_a}
\end{equation}

Finally we require that the state $\tilde{\omega}$ is a steady
state

\begin{equation}
\lim_N \tilde{\omega}([H_N,X])=0
\end{equation}
Due to the product structure of the state, the Hamiltonian can be
identified with an effective Hamiltonian \cite{fannes:1980} of the
type $\tilde{H}_N=\sum_{x\in I_N\cup II_N}\tilde{h}_x$ , where
$\tilde{h}_x\in M_{2,x}$. Imposing this time invariance yields

\begin{equation}\label{htilde}
\tilde{h}_i = \left \{
 \begin{array}{ll}
 \ei\si^z(i) -\ot(\sigma^+_{I_a})\sigma_{I_a}^-(i) -
 \ot(\sigma^-_{I_a})\sigma_{I_a}^+(i),
 & i \in I_a;\\[0.3cm]
 \ei\sigma_{I_b}^z(i) -\ot(\sigma^+_{I_a})\sigma_{I_b}^-(i) - \ot(\sigma^-_{I_a})\sigma_{I_b}^+(i)
 & \\ \qquad\qquad\qquad
 -\gamma \Big( \ot(\sigma^+_{II_b})\sigma^-_{I_b}(i) + \ot(\sigma^-_{II_b})\sigma^+_{I_b}(i)
 \Big),& i \in I_b;\\[0.3cm]
 \eii\sigma_{II_b}^z(i) -\ot(\sigma^+_{II_a})\sigma_{II_b}^-(i) -
 \ot(\sigma^-_{II_a})\sigma_{II_b}^+(i)
 & \\ \qquad\qquad\qquad
 -\gamma \Big( \ot(\sigma^+_{I_b})\sigma^-_{II_b}(i) + \ot(\sigma^-_{I_b})
 \sigma^+_{II_b}(i)\Big),& i \in II_b;\\[0.3cm]
 \eii\sigma_{II_b}^z(i) -\ot(\sigma^+_{II_a})\sigma_{II_b}^-(i) -
 \ot(\sigma^-_{II_a})\sigma_{II_b}^+(i),
  &i\in II_a,
 \end{array} \right .
\end{equation}
We use also the notation
\begin{equation}
\tilde{\Lambda}_I=\tilde{\lambda}_I\exp
i\tilde{\varphi}_I=\tilde{\omega}(\sigma_{I_b}^+), \quad
\tilde{\varphi_I}=\arg\tilde{\omega}(\sigma_{I_b}^+)
\end{equation}
and analogously for the second superconductor with $I$ replaced by $II$.\\
The local density matrices $\tilde{\rho}_x$ for $x\in {I\cup II}$
of the state $\tilde{\omega}$ are the $\tilde{h}_x$ invariant
projections of the density matrices $\rho_x$ of $\omega$
(\ref{frstate}). For more details of the construction, see
\cite{lauwers:2004}. In any case, for $x\in {I_a\cup II_a}$ :
$\rho_x=\tilde{\rho}_x$ as follows from \ref{htilde}. For the
lattice point $x\in {I_b}\cup {II_b}$ one readily computes the
selfconsistency non-equilibrium equations
\begin{equation} \label{selfcneq}
\begin{split}
\tilde{\lambda}_I& e^{i \tilde{\phi}_I}  = (\lambda_I e^{i \phi_I}
+ \gamma \tilde{\lambda}_{II} e^{i \tilde{\phi}_{II}}) \frac{
\epsilon_I^2 + \lambda_I | \lambda_I e^{i \phi_I} + \gamma
\tilde{\lambda}_{II} e^{i \tilde{\phi}_{II}} | \cos (\phi_I -
\tilde{\phi}_{I})}{\epsilon_I^2 +| \lambda_I e^{\phi_I} + \gamma
\tilde{\lambda}_{II} e^{i \tilde{\phi}_{II}}|^2} \\
\tilde{\lambda}_{II}& e^{i \tilde{\phi}_{II}} = (\lambda_{II} e^{i
\phi_{II}} + \gamma \tilde{\lambda}_{I} e^{i \tilde{\phi}_{I}})
\frac{ \epsilon_{II}^2 + \lambda_{II} | \lambda_{II} e^{i
\phi_{II}} + \gamma \tilde{\lambda}_{I} e^{i \tilde{\phi}_{I}} |
\cos (\phi_{II} - \tilde{\phi}_{II})}{\epsilon_{II}^2 +|
\lambda_{II} e^{\phi_{II}} +
\gamma \tilde{\lambda}_{I} e^{i \tilde{\phi}_{I}}|^2} \\
\end{split}
\end{equation}
Together with the selfconsistency equations \ref{lambda} for
$\lambda_I$ and $\lambda_{II}$, the equations \ref{selfcneq} form
a set of six coupled equations whose solutions determine the
non-equilibrium steady state $\tilde{\omega}$ of the total system.\\
This state divides the system into four parts. The bulk parts of
both superconductors away from the contact surface do not feel
each other nor do they feel the surface. They behave as stable
reservoirs. On the contact surfaces $I_b$ and $II_b$ the system is
effectively perturbed and influenced by the properties of the
states of both superconductors. \\ Remark that we limited the
interaction to take place only on a contact surface of one layer
thickness. It is clear that the whole construction can be
generalized to the case of any fixed finite number of layers.\\ In
\cite{lauwers:2004} we considered the currents of Cooper pairs
emerging in the system traveling from the superconductor I to II
and vice versa. One considers the relative particle number
operator of the Cooper pairs

\begin{equation}\label{numbop}
Q_N=\sum_{x\in {N^2}}(\sigma_I^+\sigma_I^-
-\sigma_{II}^+\sigma_{II}^-)
\end{equation}
The local relative current is then
\begin{equation}\label{loccurr}
J(Q_N)=i[H_N,Q_N]=-\frac{2i\gamma}{N}\sum_{i,j}^N
(\sigma_{I_b}^-(i,1)\sigma_{II_b}^+(j,1) - h.c.)
\end{equation}

Remark that in the observable current there is no direct
contribution from the bulk of the two superconductors, only the
two contact layers are contributing. Moreover one remarks that
this current is of the same order of magnitude as the contact
surface, namely $N$.\\The Josephson current measured in the
thermodynamic limit ($N\rightarrow\infty$) state $\tilde{\omega}$,
called NESS, is readily calculated and given by

\begin{equation}\label{curr}
j(Q)=\lim_N
\frac{\tilde{\omega}(J(Q_N))}{N}=-4\gamma\tilde{\lambda}_I\tilde{\lambda}_{II}\sin{(\tilde{\varphi_I}-\tilde{\varphi}_{II})}
\end{equation}

This result was obtained in \cite{lauwers:2004}, here it is
written in a more concise form. However one has to realize that
the quantities $\tilde{\lambda}$ and $\tilde{\varphi}$ are
functions of the originally given parameters
$\lambda_I$,$\lambda_{II}$,$\varphi_I$ and $\varphi_{II}$, given
by \ref{selfcneq}. First of all it is easy to see that these
equations are shift invariant for an arbitrary shift of the two
originally given angles. This means that, without loss of
generality, one can take one of the angels, say $\varphi_I$, equal
to zero. Furthermore, it is natural to assume the coupling
constant $\gamma$ to be very small, i.e.
$\gamma\ll\min{(\epsilon_I,\epsilon_{II})}$. Therefore it is
reasonable to compute the quantities $\tilde{\lambda}$ and
$\tilde{\varphi}$ only up to first order in this parameter
$\gamma$.\\ By multiplying the two selfconsistency equations with
each other, using the fact that $\varphi_I = 0$ and taking
$\varphi_{II}<\frac{\pi}{2}$ one gets that the phase difference
$\tilde{\varphi_I}-\tilde{\varphi_{II}}$ is proportional to the
phase difference $\varphi_I - \varphi_{II}$. Suppose now that also
the second phase $\varphi_{II}$ is small, corresponding to the
usual experimental regime. Then one remarks that also
$\tilde{\varphi_I}=0$. It follows that

\begin{equation}\label{eqphase}
\tilde{\varphi_{II}}\approx\varphi_{II}
\end{equation}
After substitution of \ref{eqphase} in \ref{selfcneq} one gets
\begin{eqnarray}\label{tildelambda}
\tilde{\lambda}_I=\lambda_I -
\gamma\frac{\lambda_I^2\lambda_{II}}{\mu_I^2}\\
\tilde{\lambda}_{II}=\lambda_{II} +
\gamma\frac{\lambda_I^2\epsilon_{II}^2}{\mu_{II}^2} \nonumber
\end{eqnarray}
After substitution of all these equations in the formula
\ref{curr} one gets the expected formula for the Josephson current

\begin{equation}
j(Q)=-4\gamma\lambda_I\lambda_{II}\sin(\varphi_I - \varphi_{II})
\end{equation}
yielding an analytical expression for the current for small phase
differences between the two bulk superconductors. A numerical
computation of the current for arbitrary phase differences is
found in \cite{lauwers:2004}.

\section{Symmetry breaking and Goldstone bosons}

As is well known, spontaneous symmetry breakdown (SSB) is one of
the basic features accompanying collective phenomena. It became a
representative tool for the analysis of many phenomena in modern
physics. For long range interactions, it is typical that SSB is
accompanied also by the breaking of the symmetry of the dynamics.
The latter phenomenon is known to be accompanied by the occurrence
of oscillations of a Goldstone boson with a non-vanishing energy
spectrum. These oscillations together with the Goldstone boson
disappear if the SSB disappears. In \cite{michoel2001} one was
able to construct explicitly the normal coordinates of these new
particles called Goldstone particles. In particular for mean field
systems such as the BCS-model \cite{goderis1991}, the Overhauser
model \cite{broidioi1991}, a spin density wave model
\cite{broidioi19912}, the anharmonic crystal model
\cite{verbeure1992}, and the jellium model \cite{broidioi1993},
one has constructed these Goldstone boson normal coordinates.\\

Our two-dimensional model consisting of two interacting
superconductors also shows the phenomenon of SSB. As the main
contribution of this paper we consider the construction and the
calculation of the spectrum of the corresponding Goldstone
bosons.\\

As far as the Josephson current, computed in the previous section,
is concerned we remark that the bulk parts $I_a$ and $II_a$ of the
superconductors are not contributing to it. Therefore it is
reasonable to look for the Goldstone particles within the contact
areas $I_b$ and $II_b$. In particular we compute the normal
coordinates and the dynamics of the Goldstone bosons of this
junction area. From \cite{michoel2001} we know that the canonical
coordinates of these bosons are given by the fluctuation operators
of the generator of the broken symmetry and of the order parameter
operator.\\

As the following gauge transformation holds

\begin{equation}
e^{i\alpha\sigma^z}\sigma^+ e^
{-i\alpha\sigma^z}=\sigma^+e^{2i\alpha}; \quad
\alpha\in{\mathds{R}}
\end{equation}
the $\sigma^z$ are the local generators of the broken gauge
symmetry of the effective Hamiltonian \ref{htilde}. Indeed for all
$\alpha$
\begin{equation}
\tilde{\omega}(e^{i\alpha\sigma^z}\sigma^+e^{-i\alpha\sigma^z})=\tilde{\omega}(\sigma^+)e^{2i\alpha}
\end{equation}
proving that the state is not invariant under the $U(1)$ gauge
group.\\
Therefore we consider the local operators, for $i\in{I_b,II_b}$
\begin{equation}\label{opQ}
\tilde{Q_i}=\frac{|\tilde{\Lambda}_i^2|}{\tilde{\mu}_i^2}\sigma_i^z
+ \frac{\epsilon_i}{\tilde{\mu}_i^2}
(\overline{\tilde{\Lambda}}_i\sigma_i^+ + h.c.)
\end{equation}
where for all $i\in I_b$:
\begin{eqnarray}
\tilde{\Lambda}_i = \tilde{\Lambda}_{I_b} =
\tilde{\omega}(\sigma_{I_b}^+) + \gamma
\tilde{\omega}(\sigma_{II_b})\\
\tilde{\mu}_i = \tilde{\mu}_{I_b} = \sqrt{{\epsilon_I}^2+
|\tilde{\Lambda}_{I_b}|^2}\\
\epsilon_I = \epsilon_i
\end{eqnarray}
and equivalently with $i\in {II_b}$ i.e. by substitution of $I$ by $II$ and vice versa.\\
Remark that the operator $\tilde{Q_i}$ is indeed the generator of
the gauge transformations, namely up to a constant equal to
$\sigma^z$, but normalized to zero expectation value
\begin{equation}
\tilde{\omega}(\tilde{Q}_i) =
\frac{|\tilde{\Lambda_i}^2|}{\tilde{\mu_i}^2}\tilde{\omega}(\sigma^z)
+ \frac{\epsilon_i}{\tilde{\mu_i}^2}2|\tilde{\Lambda}|^2 = 0
\end{equation}
We consider also essentially the order parameter operator
$\sigma^\pm$ fluctuation
\begin{equation}\label{opP}
\tilde{P}_j = \frac{i}{\tilde{\mu}_j}
(\overline{\tilde{\Lambda}_j}\sigma_j^+ - h.c.)
\end{equation}
Again remark that $\tilde{\omega}(\tilde{P}_j) = 0$ i.e. also this
operator is duly normalized to zero.\\

Using the general quantum fluctuation theory for product states
\cite{goderis1989}, one computes the following quantum central
limits in the given state $\tilde{\omega}$ and obtain the normal
coordinates of two Goldstone bosons. For the region $I_b$ one gets
the normal coordinates
\begin{eqnarray}\label{bosonI}
b_{I_b}(Q) = \lim_N \frac{1}{\sqrt{N}}\sum_{j\in {I_b},j=1}^N
\tilde{Q_j}\\
b_{I_b}(P) =\lim_N \frac{1}{\sqrt{N}}\sum_{j\in {I_b},j=1}^N
\tilde{P_j} \nonumber
\end{eqnarray}
and for the region ${II}_b$ one gets the normal coordinates
\begin{eqnarray}\label{bosonII}
b_{{II}_b}(Q) &= \lim_N \frac{1}{\sqrt{N}}\sum_{j\in
{{II}_b},j=1}^N \tilde{Q_j}\\
b_{{II}_b}(P) &=\lim_N \frac{1}{\sqrt{N}}\sum_{j\in
{{II}_b},j=1}^N \tilde{P_j} \nonumber
\end{eqnarray}
In \ref{bosonI} one gets the normal coordinates of a first
Goldstone boson, and in \ref{bosonII} the normal coordinates of a
second independent Goldstone boson. Indeed, by a straightforward
computation one checks readily the following canonical commutation
relations
\begin{eqnarray} \label{ccr}
&[b_{I_b}(Q),b_{{II}_b}(Q)]  = [b_{I_b}(Q),b_{{II}_b}(P)] = \\
&[b_{I_b}(P),b_{{II}_b}(Q)] = [b_{I_b}(P),b_{{II}_b}(P)] = 0  \nonumber \\
&[b_{I_b}(Q),b_{I_b}(P)] = 4i\frac{\tilde{\lambda}_{I_b}^2}{\tilde{\mu}_{I_b}} \nonumber \\
&[b_{{II}_b}(Q),b_{{II_b}}(P)]
=4i\frac{\tilde{\lambda}_{{II}_b}^2}{\tilde{\mu}_{{II}_b}}
\nonumber
\end{eqnarray}
Remark that in the case of temperatures above the critical ones of
the bulk superconductors the order parameters vanish:
$\tilde{\lambda}_{I_b} = \tilde{\lambda}_{{II}_b} = 0$ such that
all commutators in \ref{ccr} vanish. Also one has
\begin{equation}
\tilde{\omega}(b_{I_b}(Q)^2) = \tilde{\omega}(b_{{II}_b}(Q)^2) =
\tilde{\omega}(b_{I_b}(P)^2) = \tilde{\omega}(b_{{II}_b}(P)^2) = 0
\end{equation}
and hence all the operators themselves vanish: $b_{I_b}(Q) =
b_{I_b}(P) =b_{{II}_b}(Q) = b_{{II}_b}(Q) = 0 $, i.e. the
Goldstone bosons
disappear in the normal phases.\\
Next we consider the dynamics of the Goldstone bosons in the case
of superconducting phases for the bulk superconductors. We
consider the time evolution of the normal modes \ref{opQ} and
\ref{opP} which is induced by the initial micro-dynamics given by
the effective Hamiltonian \ref{htilde}. In general, let $A$ be a
local observable situated at the lattice point $x\in I_b$ or
$II_b$, then denote $\tilde{\alpha_t}$ the time evolution of the
fluctuation of A after time t. It is given by
\begin{eqnarray}\label{dyn}
\tilde{\alpha}_t b_{I_b}(A)=
b_{I_b}(\exp(it\tilde{h}_x)A\exp(-it\tilde{h}_x))\\
\tilde{\alpha}_t b_{{II}_b}(A)=
b_{{II}_b}(\exp(it\tilde{h}_x)A\exp(-it\tilde{h}_x)) \nonumber
\end{eqnarray}
Of course the operator $A$ stands for the operators $\tilde{Q}_x$
(\ref{opQ}) and $\tilde{P}_x$ (\ref{opP}).\\
A straightforward computation of the dynamics using \ref{dyn}
yields the simple solutions
\begin{eqnarray}
\tilde{\alpha_t}b_{I_b}(Q)= b_{I_b}(Q)\cos(2\tilde{\mu}_{I_b}t) +
b_{I_b}(P)\sin(2\tilde{\mu}_{I_b}t)\\
\tilde{\alpha}_t b_{I_b}(P)= - b_{I_b}(Q)\sin(2\tilde{\mu}_{I_b}t)
+ b_{I_b}(P)\cos(2\tilde{\mu}_{I_b}t) \nonumber
\end{eqnarray}
and analogously for the surface ${II}_b$ one gets the same
dynamics for the second Goldstone boson by replacing the index
$I_b$ by the index $II_b$.\\
The two bosons behave dynamically as two independent quantum
harmonic oscillators with frequencies
\begin{eqnarray}\label{freq}
\tilde{\nu}_I = 2\tilde{\mu}_{I_b}\\
 \tilde{\nu}_{II} = 2\tilde{\mu}_{{II}_b} \nonumber
\end{eqnarray}
For the bulk superconductors $I_a$ and ${II}_a$ the Goldstone
bosons dynamics were computed before \cite{goderis1991}. The
frequencies $\nu_I = 2\mu_I$ and $\nu_{II} = 2\mu_{II}$ are
clearly phase independent. \\
However for the frequencies of the Goldstone bosons considered in
this paper, the situation is completely different. The frequencies
\ref{freq} computed above do depend on the phase difference
$\varphi_I - \varphi_{II}$ of the two phases of the bulk
superconductors.\\

We consider the phase dependence explicitly for the region $I_b$,
the computation for the second region is analogous:
\begin{equation}\label{phasedep}
\tilde{\nu}_I = 2\sqrt{\epsilon_I^2+|\tilde{\Lambda_{I_b}}|^2} =
2\sqrt{\epsilon_I^2 + |\lambda_I\exp{i\varphi_I} +
\gamma\tilde{\lambda}_{II}\exp i\tilde{\varphi}_{II}|^2}
\end{equation}
Remark that the $\tilde{\lambda}_I$ and $\tilde{\varphi}_{II}$ are
determined by the parameters $\epsilon_I$, $\epsilon_{II}$,
$\lambda_I$, $\lambda_{II}$ and the phases $\varphi_I$ and
$\varphi_{II}$ through the selfconsistency equations
\ref{selfcneq}.

It is instructive again to get an explicit form of the frequencies
in terms of the given parameters of the system. We derive again
the formula for the frequency $\tilde{\nu_I}$ up to first order in
the coupling constant $\gamma$, and in the case that the phase
difference $\varphi_I - \varphi_{II}$ is small, a situation
explored before for the current.\\
We get from \ref{eqphase}, \ref{tildelambda} and \ref{freq}
\begin{equation}
\tilde{\mu}_I^2 = \epsilon_I^2 + \lambda_I^2 +
2\gamma\lambda_I\lambda_{II}\cos(\varphi_I - \varphi_{II})
\end{equation}
Hence one gets the following expressions for the dynamical
frequencies of the Goldstone modes
\begin{eqnarray}
\tilde{\nu_I} = \nu_I +
4\gamma\frac{\lambda_I\lambda_{II}}{\nu_I}\cos(\varphi_I -
\varphi_{II})\\
\tilde{\nu_{II}} = \nu_{II} +
4\gamma\frac{\lambda_I\lambda_{II}}{\nu_{II}}\cos(\varphi_I -
\varphi_{II})
\end{eqnarray}
From these expressions the dependence of the Goldstone frequencies
on the phase differences of the bulk superconductors is explicitly
given. We learn that these frequencies decrease when the phase
difference increase in contradistinction with the current. The
current has a sine-behavior, the frequencies a cosine-behavior.
This point may be interesting from the experimental point of view.

\bibliographystyle{plain}

\end{document}